# A Single-Emitter Gain Medium for Bright Coherent Radiation from a Plasmonic Nanoresonator

**Pu Zhang**[1]

[1]School of Physics, Huazhong University of Science and Technology, Luoyu Road 1037, Wuhan, 430074, People's Republic of China

Email: puzhang0702@hust.edu.cn

**Igor Protsenko**[2]

[2]Lebedev Physical Institute, Leninsky prospect 53, Moscow, 119991, Russia

Email: protsenk@gmail.com

**Vahid Sandoghdar**[3]

[3]Max Planck Institute for the Science of Light, Staudtstr. 2, D-91058 Erlangen, Germany

Email: vahid.sandoghdar@mpl.mpg.de

**Xue-Wen Chen**[1]*

**\*Corresponding author**

[1]School of Physics, Huazhong University of Science and Technology, Luoyu Road 1037, Wuhan, 430074, People's Republic of China

Email: xuewen_chen@hust.edu.cn

Telephone: +86-027-8755 7374; +86-15107153502

Fax: +86-027-8755 6576

# Abstract

We propose and demonstrate theoretically bright coherent radiation from a plasmonic nanoresonator powered by a single three-level quantum emitter. By introducing a dual-pump scheme in a Raman configuration for the three-level system, we overcome the fast decay of nanoplasmons and achieve macroscopic accumulation of nanoplasmons on the plasmonic nanoresonator for stimulated emission. We utilize the optical antenna effect for efficient radiation of the nanoplasmons and predict photon emission rates of 100 THz with up to 10 ps duration pulses and GHz repetition rates with the consideration of possible heating issue. We show that the ultrafast nature of the nanoscopic coherent source allows for operation with solid-state emitters at room temperature in the presence of fast dephasing. We provide physical interpretations of the results and discuss their realization and implications for ultra-compact integration of optoelectronics.

## INTRODUCTION

Conventional lasers have dimensions larger than or comparable to the wavelength of the emitted light and contain a large number of active emitters[1]. In recent years there has been great interest in developing subwavelength lasers both from a fundamental point of view and the perspective of device miniaturization and integration for applications[2-4]. To get around the need for an optical cavity, various nanolasers or spasers have been theoretically proposed[5-10] and experimentally demonstrated[11-17] based on plasmonic effects in one, two or three dimensions. Nevertheless, these devices require high concentrations of emitters to reach sufficient gain and to overcome both the radiation loss and the dissipation in metals, which may generate excess heat[3]. The high concentration of emitters may also cause quenching due to energy transfers among closely packed emitters[8,11]. Moreover, such spasers exhibit performance limitations predicted by theoretical studies[18-20]. In this article, we explore and devise a type of nanoscopic coherent light source with minimal amount of gain material, namely, generation and radiation of coherent nanoplasmons powered by only one quantum emitter. The nanoscopic light source has fields concentrated at deep subwavelength scale in three dimensions and radiates efficiently via the optical antenna effect with a photon emission rate of 100 Terahertz. We show a pathway to achieve that and elucidate the unique physics associated.

Optical antennas are metal nanostructures with localized surface plasmon resonances and can greatly enhance absorption and radiation of optical energy for a nearby quantum emitter[21]. They have been widely applied to modify spontaneous emission rate and emission pattern of single emitters[22-27]. An alternative view on optical antennas is to consider them as storage cavities with nanoscale mode volumes and low quality factors (Q) owing to high radiation loss and absorption in metal[28,29]. In spite of low Qs, optical antennas could provide high radiation efficiency[23,26] and in the meanwhile ultrastrong enhancement of the local density of photonic states (LDOS), and may lead to the so-called strong coupling between a single emitter and the nanocavity mode[30-34]. Indeed, recent experiments reported the strong coupling with vacuum Rabi splitting (VRS) in the order of 100 millielectronvolts (meV) (24.2 THz) for a single emitter[35,36]. In the emitter-antenna systems with large LDOS enhancement, we explore the possibility of obtaining bright coherent emission from a single solid-state emitter at room temperatures.

**MATERIALS AND METHODS**

Nanoplasmons on optical antennas typically have relaxation times of below 20 fs. Such fast decay not only poses a challenge for achieving strong coupling but also presents difficulties in effective pumping of the emitter. Indeed, the (non-)radiative relaxation rate of the excited emitter to the upper laser level is usually a few orders of magnitude slower than the nanoplasmon decay rate. Therefore, the average nanoplasmon number on the antenna is far less than one even when the emitter and nanoplasmons are strongly coupled. This is in stark contrast to single-atom microlasers[37-43], where the emitter-cavity coupling rates are in the range of MHz to GHz and therefore population inversion can be obtained via incoherent pumping. In order to get around the hurdle of ineffective pumping for our system, we propose to use a three-level emitter and apply a Raman pump scheme.[44] Figure 1 shows the level scheme of the emitter. The transition $|2\rangle \leftrightarrow |1\rangle$ resonantly couples to the nanoplasmons on an optical antenna (see inset). Emitters with a similar level scheme have been studied in different contexts[45,46], and such level scheme with optical dipolar transitions can be realized with emitters that do not have inversion symmetry[47], such as picene[48] and chiral fullerene $C_{76}$[49]. In contrast to conventional lasers with one pumping channel $\Omega_{13}$, we introduce a second pump field $\Omega_{23}$ so that the population in $|3\rangle$ can be rapidly transferred to $|2\rangle$, thus ensuring its occupation in the presence of fast coupling to the nanoplasmons. The depletion of the occupancy at $|3\rangle$ by $\Omega_{23}$ resembles in some sense stimulated-emission depletion (STED)[50], but here it is depleted to the lasing level $|2\rangle$ instead of the ground state. In this way, as the coupling between transition $|2\rangle \leftrightarrow |1\rangle$ and the nanoplasmons becomes strong, the occupation at $|2\rangle$ leads to macroscopic accumulation of nanoplasmons on the antenna and thereby considerable stimulated emission and lasing-like features, as discussed below.

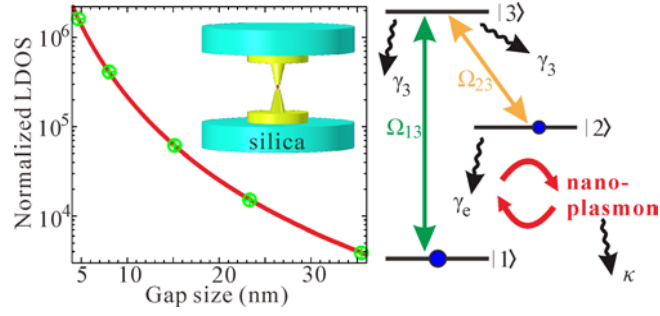

**Figure 1** The schematic of the nanoscopic coherent light source composed of a single three-level quantum emitter and an optical antenna consisting of gold double nanocones with gold nanodisk bases sitting on the silica substrates. The nanocones have 20 nm base diameter, 45 nm cone length and 6 nm flat tip diameter. The gold nanodisk is 250 nm in diameter and 50 nm in thickness. The five markers on the LDOS trace from right to left correspond to coupling constant g = 5, 10, 20, 50 and 100 meV.

For a quantitative discussion, we consider in Figure 1 an emitter coupled to a bowtie antenna structure consisting of two nanocones with gold nanodisk bases sitting on the silica substrates, providing simultaneous deep subwavelength field confinement, efficient radiation[51] and good heat dissipation. The main function of the metal disk is to increase the total volume of the metal antenna and consequently the heat capacity to circumvent the heating problem due to the absorption of the pump fields and the emission. The optical properties of the antenna (including the spectra of LDOS and absorption cross section) have been obtained through the body of revolution finite-difference time-domain (FDTD) simulations,[52,53] where the tabulated dielectric constant of gold[54] and the refractive index of 1.5 for silica are used (see

Supporting Information for more details). For the parameters used in Figure 1, the antenna has a plasmon resonance around 1.24 eV (1.0 μm wavelength, 300 THz, slightly depending on the gap size) that can be fitted with a Lorentzian line with a full width at half maximum of $\kappa = 40$ meV (9.67 THz) or a quality factor of 31. The decay rate $\kappa$ includes both the nonradiative decay caused by dissipation in the metal and the radiative decay due to the radiation of photons to the far field. The radiation efficiency of the dipolar nanoplasmon mode of the nanocavity, i.e., the ratio of the radiative decay rate to the total decay rate, is $\eta=53\%$. The trace on the left panel of Figure 1 plots $\rho_L$ the LDOS at the gap center (emitter position) normalized by the vacuum value as a function of the gap size. The coupling between the nanoplasmons and the emitter is characterized by a coupling constant g related to $\rho_L$ via[31,55] $g^2 = \kappa \gamma_e (\rho_L - 1)/4$. Here the spectral response of the antenna system has a Lorentzian line shape and the emitter-plasmon coupling constant can be evaluated through the above formula without the need of computing the effective volume of the lossy plasmon mode.[29,55] The free-space spontaneous emission rate $\gamma_e$ of $|2\rangle$ is taken to be 1 GHz (1 ns lifetime) for convenience and an emitter with a different rate leads to a different coupling constant, which will be studied later. Throughout the paper, we have assumed an ideal quantum yield for transition $|2\rangle$ to $|1\rangle$ and for a non-ideal quantum yield the radiative part of the spontaneous emission rate should be used for evaluating the coupling constant. The emitter-antenna coupled system under pumps can be treated quantum-mechanically by quantizing the nanoplasmons in the same way as photons in a cavity[32,56]. We consider a Hilbert space expanded by the emitter subspace and the Fock states of the nanoplasmons. The joint density operator $\hat{\rho}$ is a tensor product of the density matrices of the two sub-systems and evolves in time according to the master equation in Lindblad form[56]:

$$\dot{\hat{\rho}} = \frac{i}{\hbar}[\hat{\rho}, \mathcal{H}] + \mathcal{L}\left\{\sqrt{\gamma_e}\hat{\sigma}_{12}\right\} + \mathcal{L}\left\{\sqrt{\gamma_3}\hat{\sigma}_{13}\right\} + \mathcal{L}\left\{\sqrt{\gamma_{de}}\hat{\sigma}_{22}\right\} + \mathcal{L}\left\{\sqrt{\kappa}\hat{a}\right\} \qquad (1)$$

where the system Hamiltonian is time dependent and reads

$$\mathcal{H} = \hbar\omega_{21}\hat{\sigma}_{12}^\dagger\hat{\sigma}_{12} + \hbar\omega_{31}\hat{\sigma}_{13}^\dagger\hat{\sigma}_{13} + \hbar\omega_{21}\hat{a}^\dagger\hat{a} + \hbar g\left(\hat{a}+\hat{a}^\dagger\right)\left(\hat{\sigma}_{12}+\hat{\sigma}_{12}^\dagger\right)$$
$$+\hbar\Omega_{13}\cos(\omega_{31}t)\left(\hat{\sigma}_{13}+\hat{\sigma}_{13}^\dagger\right) + \hbar\Omega_{23}\cos(\omega_{32}t)\left(\hat{\sigma}_{23}+\hat{\sigma}_{23}^\dagger\right) \quad (2)$$

and the Lindblad superoperator which accounts for the various dissipations is defined as

$$\mathcal{L}\{\hat{O}\} = \hat{O}\hat{\rho}\hat{O}^\dagger - \tfrac{1}{2}\left(\hat{O}^\dagger\hat{O}\hat{\rho} + \hat{\rho}\hat{O}^\dagger\hat{O}\right). \quad (3)$$

Here $\hat{\sigma}_{ij}$ is the atomic operator $|i\rangle\langle j|$ and $\omega_{ij}$ is the transition frequency of $|i\rangle \to |j\rangle$. $\hat{a}$ and $\hat{a}^\dagger$ are the annihilation and creation Bose-operators for the nanoplasmons. We consider here that the emitter is resonantly pumped by two coherent fields with respective Rabi frequencies of $\Omega_{13}$ and $\Omega_{23}$, although in principle effective pumping can be achieved with pumps that have detunings[44]. The possible coupling of the non-resonant transitions with the pump fields can be avoided by properly arranging their polarization states if the transition dipoles orient along different directions or minimized by adjusting the pump rates according to the magnitude of dipole moments and the detuning of the transitions. As illustrated by Eq. (2), we go beyond the commonly used rotating wave approximation (RWA) to correctly describe the unique property of the strongly coupled emitter-nanocavity system. Our system shows ultrafast dynamics due to the fact that the emitter-nanocavity coupling constant g and the required pump rates are pretty large and the latter may become comparable to the optical transition frequencies. Therefore it is essential to consider the full time-dependent Hamiltonian to properly treat the fast dynamics. We remark that this treatment is very different from the modeling of microlasers with a single atom[37,40] or quantum dot[41,42], where the coupling constant and pumping rates are typically at least 5 orders of magnitude smaller than the optical transitions.

We assume that state $|3\rangle$ has a relaxation rate of $\gamma_3$=1 GHz (1 ns lifetime) to $|2\rangle$ and $|1\rangle$. The spontaneous emission rates of transition $|3\rangle \to |1\rangle$ and $|3\rangle \to |2\rangle$ do not play important roles due to the existence of the large pump fields. Additional dephasing of state $|2\rangle$ may be present with a rate $\gamma_{de}$. As an example, we take $\hbar\omega_{31}$ =3.0 eV (725 THz) and $\hbar\omega_{21}$ =1.24 eV (300 THz). Eq.(1) can be solved numerically by truncating the Fock state subspace of the nanoplasmon and propagating in time. By tracing out the atomic part of the density matrix, we can calculate the occupation of the nanoplasmon at various Fock states and find the average nanoplasmon number as $n_{NP} = \text{Tr}(\hat{a}^\dagger \hat{a} \hat{\rho})$. The second-order intensity correlation function can be evaluated according to $g^{(2)}(\tau) = \frac{\langle \hat{a}^\dagger(0)\hat{a}^\dagger(\tau)\hat{a}(\tau)\hat{a}(0)\rangle}{\langle \hat{a}^\dagger(0)\hat{a}(0)\rangle^2}$, while the power spectrum of the nanoplasmon is known as the Fourier transform of the first-order correlation $S(\omega) = \int_{-\infty}^{\infty} \langle \hat{a}^\dagger(\tau)\hat{a}(0)\rangle e^{-i\omega\tau} d\tau$, computed through time evolution of the master equation according to the quantum regression theorem[57]. We note that due to the presence of counter-rotating terms in Eq.(2), the solutions of Eq.(1) evolve with time in term of a steady-state value plus fast oscillations with a small amplitude (~1-2% of the steady-state value), which can be averaged out as shown in Figure S3. One should note that rigorously speaking when the RWA is lifted, the Fock states are no longer the eigenstates of the system. The uses of the bare system operators in the Lindblad and the calculations of spectrum and $g^{(2)}(\tau)$ are approximations valid for systems not in ultrastrong coupling regime. The small-amplitude but fast oscillations at long time (e.g. observed in Figure S3) may be also to some extent related to this approximation. A more rigorous treatment[58] should be applied to amend the conceptual inconsistency.

**RESULTS AND DISCUSSION**

Inspired by recent experiments[35,36], we first investigate the emission properties of the emitter-antenna system with g=100 meV (24.2 THz), corresponding to a 5 nm gap in our optical antenna. Figure 2a displays $n_{NP}$ as a function of the pump rates expressed in terms of $\kappa$. One clearly observes that with only one pump turned on $n_{NP}$ is essentially zero due to the fast decay of nanoplasmons. As both pumps are turned on, $n_{NP}$ quickly increases and shows a slightly asymmetric dependence on the two pump rates indicated by the white-dashed line $\Omega_{23}=1.5\Omega_{13}$ due to the strong coupling of transition $|2\rangle \leftrightarrow |1\rangle$ and the nanocavity (the asymmetry becomes smaller as the coupling constant g decreases). By increasing the pump further, $n_{NP}$ reaches to a maximum value of 3.5 and then decreases due to the fact that the effectiveness of pumping to level $|2\rangle$ is hampered by the counter-rotating coupling at large pump rates. Figure 2b shows the spectral evolution of the nanoplasmons as a function of the pump along the dashed line by three panels. As the pump increases, the three panels show four distinct emission dynamics from the bottom to the top. (i) At the lower panel, the emission from $|2\rangle \to |1\rangle$ exhibits VRS feature as predicted by the Jaynes-Cummings model[56]. (ii) As the pump increases a bit, the Raman pump induced coherence $\hat{\sigma}_{12}$ emerges and then dominates over VRS, and emission spectra display features like a Mollow triplet[56]. The change is clearly seen around $\Omega_{13}=0.1\kappa$. The appearance of the Mollow triplet can be explained through the dressed-state picture. The pump field $\Omega_{13}$ ($\Omega_{23}$) dresses the levels of $|1\rangle$ and $|3\rangle$ ($|2\rangle$ and $|3\rangle$) and the caused splittings of $|1\rangle$ and $|2\rangle$ comprise four possible transitions: two of them have the same energy as the bare $|1\rangle \leftrightarrow |2\rangle$ transition and the other two contribute to the subsidiary peaks. (iii) Displayed in the middle panel, the Mollow triplet feature gradually disappears as the pump increases, and a single peak at the center persists and broadens. The coherence built by the Raman pumps becomes weak as compared to the coherence due to the interaction with the nanoplasmons whose number increases. (iv) As nanoplasmons accumulate, stimulated emission dominates and a prominent line narrowing feature occurs. The linewidth approximately follows the Schawlow-Townes inverse proportion with the nanoplasmon number[1]. After reaching the maximum $n_{NP}$, the spectrum broadens due to the counter-rotating coupling at large pump rates. We observe a minimal linewidth of $0.026\kappa$ at $\Omega_{13}=20\kappa$.

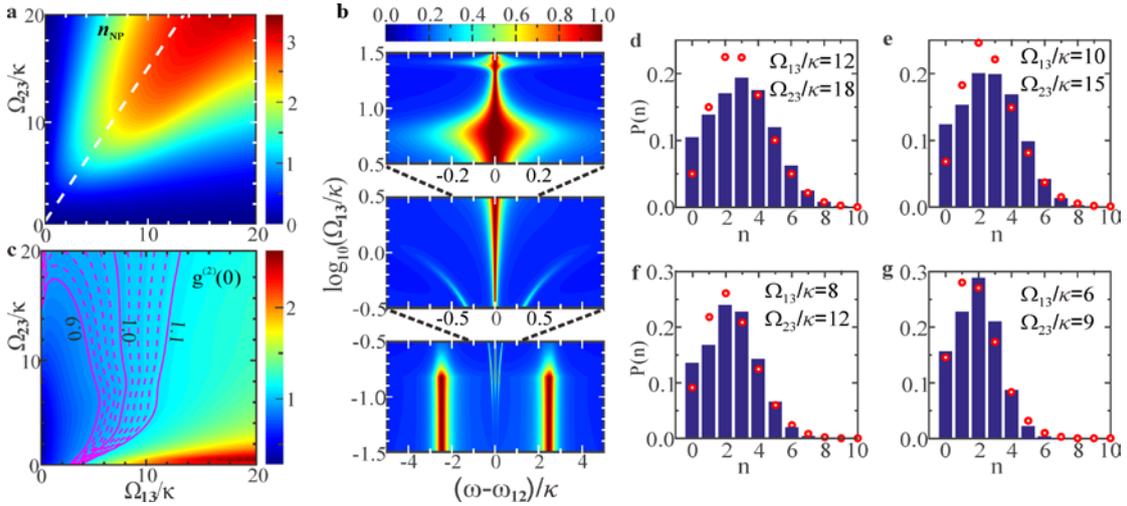

**Figure 2** Nanoplasmon generation and radiation by a single emitter with g = 100 meV. (**a**) The average nanoplasmon number vs. pump rates (dashed line: $\Omega_{23}=1.5\Omega_{13}$); (**b**) The normalized spectra of the nanoplasmon as a function of the pump rates $\Omega_{13}$ along the dashed line; (**c**) Second-order intensity correlation function at zero delay vs. pump rates; (**d**)-(**g**) Fock state occupation histograms of the nanoplasmon for various pump rates. The corresponding Poissonian distributions are shown in circles.

Figure 2c depicts the second-order intensity correlation functions at zero delay with the contours showing a range for $g^{(2)}(0)\sim 1.0$. As $\Omega_{23}$ increases and $\Omega_{13}$ is set to suitable values, the region of $g^{(2)}(0)\sim 1.0$ shows up, indicating coherent nanoplasmon generation. Additionally, the time delayed second-order intensity correlation functions $g^{(2)}(\tau)$ for four pumping parameters are presented in the Supporting Information. To investigate the coherence property of the nanoplasmon in more detail, we plot in Figure 2d-g the nanoplasmon number occupation histogram as a function of the index of the Fock state for four pumping parameters chosen along the dashed line. The corresponding Poissonian distribution with the same mean nanoplasmon number is shown by the circles. As one sees from the histogram plots, the photon statistics of our device at these pump rates are all pretty close to the Poissonian distributions, featuring similar statistics like a conventional laser with a small portion of thermal emission at low mean photon number. Thus the device can work well for a wide range of pumping parameters. For pumping rates of $\Omega_{13}=4\kappa$, $\Omega_{23}=6\kappa$, a mean nanoplasmon number of 1.2 is achieved while for $\Omega_{13}=12\kappa$, $\Omega_{23}=18\kappa$ one obtains $n_{NP}=3.0$. The pump rate of $\Omega_{23}=18\kappa$ corresponds to light intensities about several tens of GW/cm$^2$, which is readily available with pulse lasers. We remark in passing that although $n_{NP}=3.0$ might seem a moderate nanoplasmon number, the device in fact delivers coherent photon streams at a remarkable rate of $n_{NP}\kappa\eta$~100 THz (equivalent continuous-wave power of 19.4 μW) to the far field. We note that the emission rate of microlasers is typically in the order of a few Megahertz for atoms[40] or GHz for semiconductor quantum dots[42]. However, one should be cautious about the possible heating problem of the plasmonic antenna device[59,60] due to absorption of the pump fields and the emission. The present antenna structure has been designed to alleviate the heating issue and allows the device to be operated in pulse mode with pulse duration up to 10 ps. The absorption cross section spectrum and the details for the estimation of the upper limit of the temperature rise[59,60] are provided in the Supporting Information. For each excitation pulse, our single-emitter device can emit c.a. 1000 photons. The cool-down time of the device should be in the order of 100 ps to 1ns,[60] which translates to a repetition rate of 1GHz to 10 GHz. The efficiency of the single-emitter nanoscopic source depends on a number of factors, such as the dipole moment of the emitter at the pumping transitions, the field enhancement factors at the pumping frequencies and the average plasmon number to achieve. We estimate that the efficiency is on the order of 10$^{-5}$ for the present single-emitter device.

Next we investigate in more detail the properties of the single-emitter nanoscopic light source and gain insight into the working principle. Figure 3a and 3b respectively plot with color-coded traces $n_{NP}$ and state $|2\rangle$ population $\langle\hat{\sigma}_{22}\rangle$ of the emitter for g=100 meV at pumps along the dashed line ($\Omega_{23}=1.5\Omega_{13}=1.5\Omega$) in Figure 2a and for smaller g at equal pump $\Omega_{23}=\Omega_{13}=\Omega$. The populations of the atomic states are obtained from the diagonal elements of the reduced density matrix for the atomic part. For fixed g, the trends of $n_{NP}$ and $\langle\hat{\sigma}_{22}\rangle$ versus the pump are similar. In particular, for g=100 meV, the two blue-solid traces show a clear change of behavior around $\Omega=6\kappa$, which corresponds to the transition point in the spectrum evolution chart in Figure 2b. Before the transition $n_{NP}$ increases with the pump rate at a large slope due to the fast increase of $\langle\hat{\sigma}_{22}\rangle$. After the transition, $n_{NP}$ grows at a slower rate before reaching the saturation. We observe that the populations of $|2\rangle$ and $|1\rangle$ are not inverted (Figure S5) and $n_{NP}$ is mostly related to $\langle\hat{\sigma}_{22}\rangle$, and population inversion is unnecessary for the accumulation of nanoplasmons. This is because when the inversion is negative the Raman pump scheme forms an efficient path ($|1\rangle \rightarrow |3\rangle \rightarrow |2\rangle$) to cycle the population from $|1\rangle$ to $|2\rangle$, and can be much faster than the direct simulated absorption ($|1\rangle \rightarrow |2\rangle$) by the nanoplasmons. The larger $\langle \hat{\sigma}_{22} \rangle$ is, the more the plasmons can accumulate. The strong emitter-plasmon coupling basically has two effects, i.e., (1) make the emitter relax to the ground state as fast as possible for the pump to work effectively; (2) help accumulate plasmons in the nanocavity. Thanks to the near-unity coupling of the emission to the antenna mode and the pump scheme, we observe thresholdless behavior for the emission, which allows a wide range of the pumping parameters for operation as shown in Figures 2d-g. The above features are present for all coupling constants considered here. For smaller g, at large pump $\langle \hat{\sigma}_{22} \rangle$ saturates to 0.5 and $n_{NP}$ asymptotically reaches maximum values before the effect from counter-rotating coupling sets in. In the weak coupling regime, the behavior of $n_{NP}$ could be qualitatively predicted with a simple and physically more intuitive rate equation model as formulated in the Supporting Information.

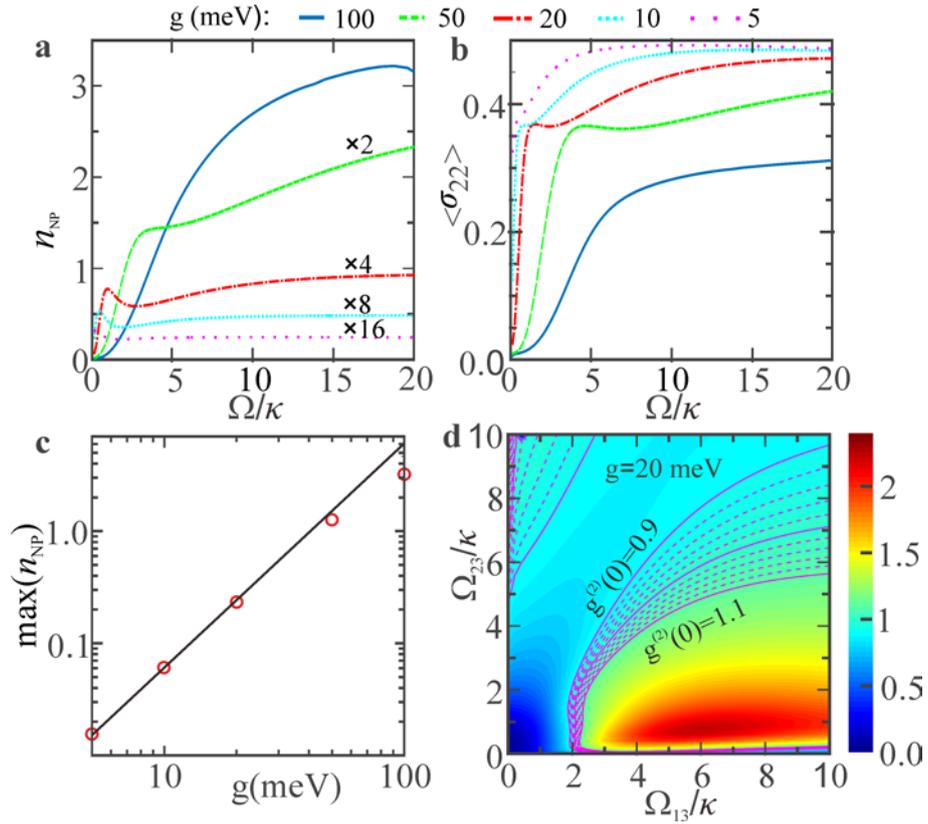

**Figure 3** Average nanoplasmon number (**a**) and population in state $|2\rangle$ (**b**) as functions of the pump rate $\Omega$ and the coupling constant. (**c**) The maximum nanoplasmon number vs. coupling constant in double logarithmic scale. (**d**) Second-order intensity correlation function at zero delay $g^{(2)}(0)$ for the coupling constant of g = 20 meV. The contours of $g^{(2)}(0)$ = 0.9 to 1.1 are overlaid on the graph.

The saturation of the nanoplasmon number for small g with the increase of the pump is a prominent difference from the self-quenching phenomena observed in microlasers[37], where the photon number in the cavity decreases at large pump rate due to decoherence. The saturation behavior in our system can be understood from the perspective of power balance. The power feeding the whole system is given by $W = -\frac{1}{2}\text{Re}\left\{i\omega_0 \langle \vec{p}^* \cdot \vec{E} \rangle\right\}$, where $\vec{p} = \vec{\mu}\hat{\sigma}_{12}$ is the induced dipole moment of the emitter and $\vec{E}$ is the electric field operator at $\omega_0$ at the emitter position. It follows that $\langle \vec{\mu} \cdot \vec{E} \rangle \propto \sqrt{n_{NP}} g$ and the power loss rate $\propto \hbar\omega_0 n_{NP} \kappa$. The feeding power and the loss should be balanced and consequently one obtains $n_{NP} \propto \left(\frac{g}{\kappa}\right)^2$. The single-emitter coherence $\langle \hat{\sigma}_{12} \rangle$ has a finite value and therefore there is an upper bound for the attainable plasmon number. In particular, for the system considered here, $\langle \hat{\sigma}_{12} \rangle$ saturates to a value of -0.5 for small g, and consequently the correlation of the coherence and field operators reaches a constant value. Figure 3c depicts the numerically calculated maximum $n_{NP}$ as a function of coupling constant and the data can be fitted by $n_{NP}=0.97(g/\kappa)^2$. Figure 3d displays the second-order intensity correlation function for g=20 meV. The contours on the graph show the range for $g^{(2)}(0) \sim 1.0$. One sees that for the coupling constant considered here good coherence of nanoplasmon generation can be obtained for a range of suitable pumping parameters. For instance, with g=20 meV, we obtain a photon emission rate of over 4.5 THz (equivalent continuous-wave power 0.9 μW) for $\Omega_{13}=6\kappa$, $\Omega_{23}=6\kappa$. The fact that for g = 20 meV the system delivers bright coherent photon streams shows that it can operate at various levels of coupling, including the case where the system is not strongly coupled. With state-of-the-art nanofabrication and nanocontrol techniques[35,36], such systems should be realizable in the laboratory.

We have studied so far the nanoplasmon generation and radiation with a single emitter without additional dephasing channel other than spontaneous decay, which only holds for solid-state emitters at cryogenic temperatures. At higher temperatures, due to the system-environment interactions, solid-state emitters usually experience fast dephasing. For example, organic molecules at room temperatures may have coherence times of 100 fs. We now discuss the dephasing effect on the performance of the device. We introduce the dephasing $\gamma_{de}$ to state $|2\rangle$ on top of spontaneous decay and calculate the average nanoplasmon number and second-order intensity correlation function as functions of pump rates. Dephasing rates of 10 GHz, 100 GHz, 1 THz and 10 THz have been considered for g = 100 meV and the results are summarized in Figure 4. $n_{NP}$ is displayed in color maps as functions of pump rates, while $g^{(2)}(0)$ contours with values from 0.9 to 1.1 are overlaid. As dephasing rate becomes larger, the contour enclosed region shrinks to a narrow band. Within the overlaid region, relatively large average nanoplasmon number ($n_{NP}$>1.5) with good coherence properties can always be guaranteed. For pump rates $\Omega_{13}=8\kappa$, $\Omega_{23}=7\kappa$, Figures 4**e**-**h** show that the Fock state occupation histograms resemble the Poissonian distributions for all the dephasing rates considered here. We note that due to the unique ultrafast nature of our system the nanoscopic coherent light source can tolerate fast dephasing and operate well with solid-state emitters at room temperatures in contrast to microlasers with single quantum dots where cryogenic temperature condition is necessary. The dephasing effects for the system in the weak coupling regime, e.g., g = 10 meV, have also been studied and shown in the Supporting Information.

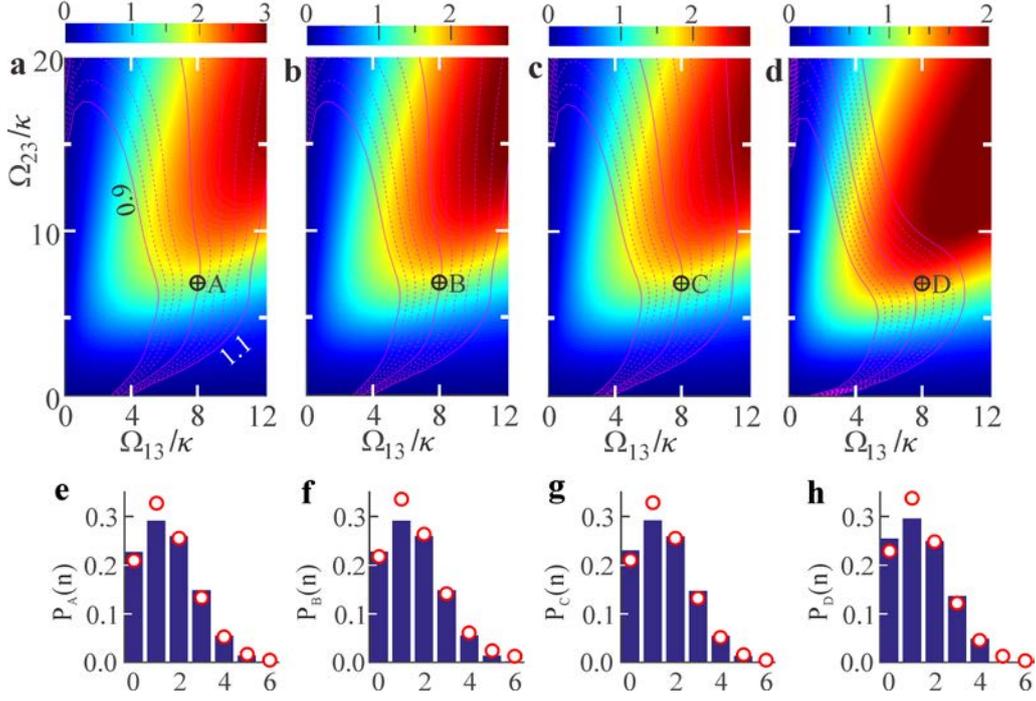

**Figure 4** Average nanoplasmon number and second-order intensity correlation function at zero delay $g^{(2)}(0)$ for g = 100 meV and dephasing rates of 10 GHz (**a**), 100 GHz (**b**), 1 THz (**c**) and 10 THz (**d**). The contours of $g^{(2)}(0)$ = 0.9 to 1.1 from left to right are overlaid on the graphs. Lower panel (**e**)-(**h**): Fock state occupation histograms of the nanoplasmons for the points marked as A-D in (**a**)-(**d**) for comparison with Poissonian distributions (circles).

**CONCLUSIONS**

In this article, we have proposed the generation and radiation of coherent nanoplasmons fueled by a single-emitter gain medium and investigated the emission properties of the device at various conditions based on a fully quantized model beyond the RWA to account for the ultrafast dynamics. The Raman pump scheme effectively expedites the process of populating the upper emitting level to overcome the fast decay of nanoplasmons and boosts the average nanoplasmon number from otherwise far less than one to a macroscopic accumulation. In the proposed system, the plasmonic nanoresonator provides deep-subwavelength field concentration and meanwhile efficiently funnels the energy to far-field photons via the optical antenna effect. With realistic parameters including the coupling constant, pump rate and dephasing rate, our calculations show that the single-emitter nanoscopic source can deliver bright coherent photon streams with a rate of 100 THz with up to 10 ps duration pulses and GHz repetition rates with the consideration of possible heating issue. The ultrafast nature of the proposed device allows to work with solid-state emitters and in pulse mode with minimum heating problem[3]. The proposed scheme bridges the size mismatch between optical components and individual emitters, making it ideally suitable for integrated plasmonic circuitry. The possibility of achieving nanoscale bright coherent emission with a single solid-state emitter has far-reaching implications for ultra-compact integration of functional optoelectronic devices[4] and for quantum plasmonics and metamaterials[61].

**AUTHOR INFORMATION**

**Corresponding Author**

*E-mail: xuewen_chen@hust.edu.cn.

**Notes**

The authors declare no competing financial interest.

**ACKNOWLEDGEMENTS**

We acknowledge financial support from the National Natural Science Foundation of China (grant 11474114, 11604109), the Thousand-Young-Talent Program of China, Huazhong University of Science and Technology, and the Max Planck Society. X.-W.C acknowledges fruitful discussions with Z.-Y. Li, R.-M. Ma, C.-F. Zhang and M. Litinskaya. I.P. wishes to thank I.V. Smetanin and A. Uskov for discussions.

**Supporting Information Available**: <Supporting Information.> This material is available free of charge via the Internet at http://pubs.acs.org. In the Supporting Information, we provide additional information regarding the electromagnetic simulations, the estimation of temperature rise, the time evolution of nanoplasmon number and the second-order intensity correlation function and a rate equation model for estimating the average plasmon number at low g-factors.

**REFERENCES**

(1) Siegman, A. E. Lasers; University Science Books: Sausalito, CA (USA), 1986.
(2) Berini, P.; De Leon, I. Surface plasmon-polariton amplifiers and lasers. *Nat. Photonics* **2012,** *6*, 16-24.
(3) Hess, O.; Pendry, J. B.; Maier, S. A.; Oulton, R. F.; Hamm. J. M.; Tsakmakidis, K. L. Active nanoplasmonic metamaterials. *Nat. Mater.* **2012,** *11*, 573-584.
(4) Hill, M. T.; Gather, M. C. Advances in small lasers. *Nat. Photonics* **2014,** *8*, 908-918.
(5) Bergman, D. J.; Stockman, M. I. Surface plasmon amplification by stimulated emission of radiation: Quantum generation of coherent surface plasmons in

nanosystems. *Phys. Rev. Lett.* **2003,** *90*, 027402-027405.

(6) Protsenko, I. E.; Uskov, A. V.; Zaimidoroga, O. A.; Samoilov, V. N.; OReilly, E. P. Dipole nanolaser. *Phys. Rev. A* **2005,** *71*, 063812-063818.

(7) Zheludev, N. I.; Prosvirnin, S. L.; Papasimakis, N.; Fedotov, V. A. Lasing spaser. *Nat. Photonics* **2008,** *2*, 351-354.

(8) Stockman, M. I. Spasers explained. *Nat. Photonics* **2008,** *2*, 327-329.

(9) Chang, S. W.; Ni, C. Y. A.; Chuang, S. L. Theory for bowtie plasmonic nanolasers. *Opt. Express* **2008,** *16*, 10580-10595.

(10) Dorfman, K. E.; Jha, P. K.; Voronine, D. V.; Genevet, P.; Capasso, F.; Scully, M. O. Quantum-coherence-enhanced surface plasmon amplification by stimulated emission of radiation. *Phys. Rev. Lett.* **2013,** *111*, 043601-043605.

(11) Noginov, M. A.; Zhu, G.; Belgrave, A. M.; Bakker, R.; Shalaev, V. M.; Narimanov, E.; Stout, S.; Herz, E.; Suteewong, T.; Wiesner, U. Demonstration of a spaser-based nanolaser. *Nature* **2009,** *460*, 1110-1113.

(12) Oulton, R. F.; Sorger, V. J.; Zentgraf, T.; Ma, R. M.; Gladden, C.; Dai, L.; Bartal, G.; Zhang, X. Plasmon lasers at deep subwavelength scale. *Nature* **2009,** *461*, 629-632.

(13) Ma, R. M.; Oulton, R. F.; Sorger, V. J.; Bartal, G.; Zhang, X. Room-temperature sub-diffraction-limited plasmon laser by total internal reflection. *Nat. Mater.* **2011,** *10*, 110-113.

(14) Lu, Y. J.; Kim, J.; Chen, H. Y.; Wu, C. H.; Dabidian, N.; Sanders, C. E.; Wang, C. Y.; Lu, M. Y.; Li, B. H.; Qiu, X.; Chang, W. H.; Chen, L. J.; Shvets, G.; Shih, C. K.; Gwo, S. Plasmonic nanolaser using epitaxially grown silver film. *Science* **2012,** *337*, 450-453.

(15) Ding, K.; Liu, Z. C.; Yin, L. J.; Hill, M. T.; Marell, M. J. H.; van Veldhoven, P. J.; Noetzel, R.; Ning, C. Z. Room-temperature continuous wave lasing in deep-subwavelength metallic cavities under electrical injection. *Phys. Rev. B* **2012,** *85*, 041301(R)-041305.

(16) Zhou, W.; Dridi, M.; Suh, J. Y.; Kim, C. H.; Co, D. T.; Wasielewski, M. R.; Schatz, G. C.; Odom, T. W. Lasing action in strongly coupled plasmonic nanocavity arrays. *Nat. Nanotech.* **2013,** *8*, 506-511.

(17) Meng, X.; Liu, J. J.; Kildishev, A. V.; Shalaev, V. M. Highly directional spaser array for the red wavelength region. *Laser Photonics Rev.* **2014,** *8*, 896-903.

(18) Arnold, N.; Hrelescu, C.; Klar, T. Minimal spaser threshold within electrodynamic framework: Shape, size and modes. *Ann. Phys. (Berlin, Ger.)* **2016**, *528*, 295-306.

(19) Kewes, G.; Herrmann, K.; Rodríguez-Oliveros, R.; Kuhlicke, A.; Benson, O.; Busch, K. Limitations of Particle-Based Spasers. *Phys. Rev. Lett.* **2017**, *118*, 237402-237407.

(20) Khurgin, J. B.; Sun, G. Comparative analysis of spasers, vertical-cavity surface-emitting lasers and surface-plasmon-emitting diodes. *Nat. Photonics* **2014**, *8*, 468−473.

(21) Agio, M.; Alu, A. Optical Antennas. New York: Cambridge University Press, 2013.
(22) Greffet, J. J. Nanoantennas for light emission. *Science* **2005,** *308*, 1561-1563.
(23) Rogobete, L.; Kaminski, F.; Agio, M.; Sandoghdar, V. Design of plasmonic nanoantennae for enhancing spontaneous emission. *Opt. Lett.* **2007,** *32*, 1623-1625.
(24) Curto, A. G.; Voipe, G.; Taminiau, T. H.; Kreuzer, M. P.; Quidant, R.; Hulst, N. F. Unidirectional emission of a quantum dot coupled to a nanoantenna. *Science* **2010,** *329*, 930-933.
(25) Gu, Y.; Wang, L. J.; Ren, P.; Zhang, J. X.; Zhang, T. C.; Martin, O. J. F.; Gong, Q. H. Surface-plasmon-induced modification on the spontaneous emission spectrum via subwavelength-confined anisotropic Purcell factor. *Nano Lett.* **2012,** *12*, 2488-2493.
(26) Chen, X.-W.; Agio, M.; Sandoghdar, V. Metallodielectric hybrid antennas for ultrastrong enhancement of spontaneous emission. *Phys. Rev. Lett.* **2012,** *108*, 233001-233005.
(27) Ren, J. J.; Gu, Y.; Zhao, D. X.; Zhang, F.; Zhang, T. C.; Gong, Q. H. Evanescent-vacuum-enhanced photon-exciton coupling and fluorescence collection. *Phys. Rev. Lett.* **2017,** *118*, 073604-073609.
(28) Agio, M. Optical antennas as nanoscale resonators. *Nanoscale* **2012,** *4*, 692-706.
(29) Sauvan, C.; Hugonin, J. P.; Maksymov, I. S.; Lalanne, P. Theory of the spontaneous optical emission of nanosize photonic and plasmon resonators. *Phys. Rev. Lett.* **2013,** *110*, 237401-237405.
(30) Trügler, A.; Hohenester, U. Strong coupling between a metallic nanoparticle and a single molecule. *Phys. Rev. B* **2008,** *77*, 115403-115408.
(31) Savasta, S.; Saija, R.; Ridolfo, A.; Stefano, O. D.; Denti, P.; Borghese, F. Nanopolaritons: Vacuum Rabi splitting with a single quantum dot in the center of a dimer nanoantenna. *ACS Nano* **2010,** *4*, 6369-6376.
(32) Ridolfo, A.; Di Stefano, O.; Fina, N.; Saija, R.; Savasta, S. Quantum plasmonics with quantum dot-metal nanoparticle molecules: Influence of the Fano effect on photon statistics. *Phys. Rev. Lett.* **2010,** *105*, 263601-263604.
(33) Manjavacas, A.; García de Abajo, F. J.; Nordlander, P. Quantum plexcitonics: Strongly interacting plasmons and excitons. *Nano Lett.* **2011,** *11*, 2318-2323.
(34) Li, R. Q.; Hernángomez-Pérez, D.; García-Vidal, F. J.; Fernández-Domínguez, A. I. Transformation optics approach to plasmon-exciton strong coupling in nanocavities. *Phys. Rev. Lett.* **2016,** *117*, 107401-107405.
(35) Chikkaraddy, R.; de Nijs, B.; Benz, F.; Barrow, S. J.; Scherman, O. A.; Rosta, E.; Demetriadou, A.; Fox, P.; Hess, O.; Baumberg, J. J. Single-molecule strong coupling at room temperature in plasmonic nanocavities. *Nature* **2016,** *535*, 127-130.
(36) Santhosh, K.; Bitton, O.; Chuntonov, L.; Haran, G. Vacuum Rabi splitting in a plasmonic cavity at the single quantum emitter limit. *Nat. Comm.* **2016,** *7*,

11823-11827.

(37) Mu, Y.; Savage, C. M. One-atom lasers. *Phys. Rev. A* **1992,** *46*, 5944-5956.

(38) Rice, P. R.; Carmichael, H. J. Photon statistics of a cavity-QED laser: A comment on the laser-phase-transition analogy. *Phys. Rev. A* **1994,** *50*, 4318-4329.

(39) Löffler, M.; Meyer, G. M.; Walther, H. Spectral properties of the one-atom laser. *Phys. Rev. A* **1997,** *55*, 3923-3930.

(40) Mckeever, J.; Boca, A.; Boozer, A. D.; Buck, J. R.; Kimble, H. J. Experimental realization of a one-atom laser in the regime of strong coupling. *Nature* **2003,** *425*, 268-271.

(41) Astafiev, O.; Inomata, K.; Niskanen, A. O.; Yamamoto, T.; Pashkin, Y. A.; Nakamura, Y.; Tsai, J. S. Single artificial-atom lasing. *Nature* **2007,** *449*, 588-590.

(42) Nomura, M.; Kumagai, N.; Iwamoto, S.; Ota, Y.; Arakawa, Y. Laser oscillation in single-quantum-dot-nanocavity system. *Nat. Phys.* **2010,** *6*, 279-283.

(43) Moelbjerg, A.; Kaer, P.; Lorke, M.; Tromborg, B.; Mørk, J. Dynamical properties of nanolasers based on few discrete emitters. *IEEE J. Sel. Top. Quantum Electron.* **2013,** *49*, 945-954.

(44) Wu, Y. effective Raman theory for a three-level atom in the Λ configuration. *Phys. Rev. A* **1996,** *54*, 1586-1592.

(45) Patterson, D.; Schnell, M.; Doyle, J. M. Enantiomer-specific detection of chiral molecules via microwave spectroscopy. *Nature* **2013**, 497, 475–477.

(46) Lobsiger, S.; Perez, C.; Evangelisti, L.; Lehmann, K. K.; Pate, B. H. Molecular Structure and Chirality Detection by Fourier Transform Microwave Spectroscopy. *J. Phys. Chem. Lett.* **2015**, 6, 196–200.

(47) Kral, P.; Shapiro, M. Cyclic Population Transfer in Quantum Systems with Broken Symmetry. *Phys. Rev. Lett.* **2001**, 87, 183002-183005.

(48) Galué, H. Á.; Oomens, J.; Buma, W. J.; Redlich, B. Electron-flux infrared response to varying π-bond topology in charged aromatic monomers. *Nat. Comm.* **2016**, 7, 12633-12644.

(49) Yabana, K.; Bertsch, G. F. Application of the time-dependent local density approximation to optical activity. *Phys. Rev. A.* **1999**, 60, 1271-1279.

(50) Hell, S. W.; Wichmann, J. Breaking the diffraction resolution limit by stimulated emission: Stimulated-emission-depletion fluorescence microscopy. *Opt. Lett.* **1994,** *19*, 780-782.

(51) Mohammadi, A.; Kaminski, F.; Sandoghdar, V.; Agio, M. Fluorescence enhancement with the optical (bi-) conical antenna. *J. Phys. Chem. C* **2010,** *114*, 7372-7377.

(52) Taflove, A.; Hagness, S. C. Computational electrodynamics: the finite-difference time-domain method, 3rd ed.; Artech: Boston, 2005.

(53) Chen, X.-W.; Sandoghdar, V.; Agio, M. Highly Efficient Interfacing of Guided Plasmons and Photons in Nanowires. *Nano Lett.* **2009**, 9, 3756–3761.

(54) Haynes, W. CRC Handbook of Chemistry and Physics, 87th ed.; Haynes, W.,

Eds.; CRC Press/Taylor and Francis Group, LLC: Boca Raton, FL, 2006.
(55) Lambropoulos, P.; Petrosyan, D. Fundamentals of Quantum Optics and Quantum Information; Springer: Berlin, 2007.
(56) Garrison, J. C.; Chiao, R. Y. Quantum Optics. Oxford: Oxford University Press, 2008.
(57) Gardiner, C. W. Quantum Noise. Berlin: Springer-Verlag, 1991.
(58) Di Stefano, O.; Stassi, R.; Garziano, L.; Kockum, A. F.; Savasta, S.; Nori, F. Feynman-diagrams approach to the quantum Rabi model for ultrastrong cavity QED: stimulated emission and reabsorption of virtual particles dressing a physical excitation. *New J. Phys.* **2017**, *19*, 053010-053028.
(59) Baffou, G.; Quidant, R. Thermo-Plasmonics: using metallic nanostructures as nano-sources of heat. *Laser Photon. Rev.* **2013**, *7*, 171–187.
(60) Baffou, G.; Quidant, R.; García de Abajo, F. J. Nanoscale Control of Optical Heating in Complex Plasmonic Systems. ACS Nano **2010**, *4*, 709–716.
(61) Tame, M. S.; McEnery, K. R; Oezdemir, S. K; Lee, J.; Maier, S. A.; Kim, M. S. Quantum plasmonics. *Nat. Phys.* **2013,** *9*, 329-340.

For Table of Contents Use Only

# A Single-Emitter Gain Medium for Bright Coherent Radiation from a Plasmonic Nanoresonator

**Pu Zhang**, **Igor Protsenko**, **Vahid Sandoghdar**, **Xue-Wen Chen**

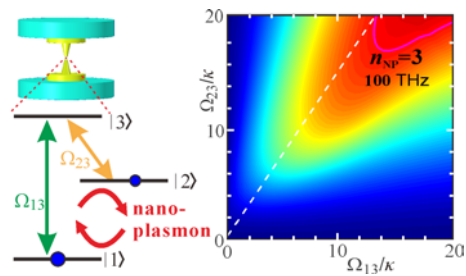

The TOC figure displays the proposed single-emitter light source consisting of coupled optical antenna and three-level system, and the dependence of coherent radiation output on the pumping rates.